\documentclass{llncs}
\usepackage{tikz}
\usepackage{hyperref}
\usepackage{amsmath}
\usepackage{amssymb}
\usepackage{listings}
\usepackage{verbatim}
\usepackage{color}
\usepackage{graphicx}
\usepackage{algorithm}% http://ctan.org/pkg/algorithm
\usepackage[noend]{algpseudocode}% http://ctan.org/pkg/algorithmicx
\usepackage{multirow}
\usepackage{pdfpages}
\usepackage{multibib}
\usepackage{wrapfig}
\usepackage{paralist}
\usepackage{afterpage}
\usepackage{subcaption}
\usepackage{booktabs}
\usepackage{footnote}
\usepackage{tablefootnote}
\usepackage{threeparttable}

\usetikzlibrary{shapes}
\usetikzlibrary{positioning}
\usetikzlibrary{arrows}
\usetikzlibrary{calc}
\usetikzlibrary{automata}
\tikzstyle{state}+=[minimum size = 6mm, inner sep=0]
\usepackage{times}
\newcites{add}{Additional References}

\newcommand{\theoremlike}[2]{\par\medskip\penalty-250\refstepcounter{theorem}{{\bfseries\noindent#2
\ref{#1}.}}}

\newcommand{\thmhelperpre}[2]{\theoremlike{#1}{#2}}
\newcommand{\thmhelperpost}{\par\medskip}

\DeclareMathOperator*{\argmax}{arg\,max}

\newcommand{\lstrat}{\varsigma}

\newcommand{\Qset}{\mathbb{Q}}
\newcommand{\Nset}{\mathbb{N}}
\newcommand{\Rset}{\mathbb{R}}
\newcommand{\Zset}{\mathbb{Z}}

\newcommand{\dist}{\mathit{Dist}}
\newcommand{\reach}{\Diamond}

\newcommand{\inits}{\hat s}
\newcommand{\act}{\mathit{Act}}
\newcommand{\pat}{\omega}
\newcommand{\Pat}{\mathsf{Runs}}
\newcommand{\fpat}{w}
\newcommand{\Cone}{\mathsf{Cone}}
\newcommand{\calF}{\mathcal{F}}
\newcommand{\mec}{\mathsf{MEC}}

\newcommand{\pr}{\mathbb P}	
\newcommand{\expected}{\mathbb E}
\newcommand{\val}{\mathit{Val}}
\newcommand{\imp}{\mathit{Imp}}

\newcommand{\true}{\mathit{good}}
\newcommand{\false}{\mathit{bad}}

\newcommand{\playaction}[1]{$action = [$\texttt{#1}$]$}
\newcommand{\thresh}{\Delta}

\usepackage{microtype}
\newcommand{\myspace}{\vspace*{-0.5em}}
\advance\textwidth4mm % <= 5mm
\advance\hoffset-2mm
\advance\textheight4mm
\advance\voffset-2mm

\title{Counterexample Explanation by Learning\\ Small Strategies in Markov Decision Processes}

\author{Tom\'a\v{s} Br\'{a}zdil$^1$ \and Krishnendu Chatterjee$^2$ \and Martin Chmel\'{i}k$^2$
\and Andreas Fellner$^2$ \and \\ Jan K\v{r}et\'insk\'y$^2$}

\institute{$^1$ Masaryk University, Brno, Czech Republic \\ $^2$ IST Austria}

\begin{document}
\maketitle

\begin{abstract}
While for deterministic systems, a counterexample to a property can simply be an error trace, counterexamples in probabilistic systems are necessarily more complex.
For instance, a set of erroneous traces with a sufficient cumulative probability mass can be used.
Since these are too large objects to understand and manipulate, compact representations such as subchains have been considered.
In the case of probabilistic systems with non-determinism, the situation is even more complex.
While a subchain for a given strategy (or scheduler, resolving non-determinism) is a straightforward choice, we take a different approach.
Instead, we focus on the strategy---which can be a counterexample to violation of or a witness of satisfaction of a property---itself, and extract the most important decisions it makes, and present its succinct representation. 

The key tools we employ to achieve this are (1) introducing a concept of importance of a state w.r.t.\ the strategy, and (2) learning using decision trees. 
There are three main consequent advantages of our approach. 
Firstly, it exploits the quantitative information on states, stressing the more important decisions.
Secondly, it leads to a greater variability and degree of freedom in representing the strategies.
Thirdly, the representation uses a self-explanatory data structure.
In summary, our approach produces more succinct and more explainable strategies, as opposed to e.g.\ binary decision diagrams.
Finally, our experimental results show that we can extract several rules describing the strategy even for very large systems that do not fit in memory, and based on the rules explain the erroneous behaviour.
\end{abstract}

%!TEX root = main.tex

\section{Introduction}

The standard models for dynamic stochastic systems with both probabilistic and nondeterministic behaviour are 
\emph{Markov decision processes} (MDPs)~\cite{Howard60,Puterman:book,FV97}.
They are widely used in verification of probabilistic systems~\cite{BK08,KNP11} in several ways.
Firstly, in concurrent probabilistic systems, such as communication protocols, the nondeterminism arises from scheduling~\cite{CY95,Var85}.
Secondly, probabilistic systems operating in open environments, such as various stochastic reactive systems, respond to nondeterministic inputs~\cite{SegalaT,dA97a}.
Thirdly, for underspecified probabilistic systems, a controller is synthesized, resolving the nondeterminism in a way that optimizes some objective, such as energy consumption or time constraints in embedded systems~\cite{BK08,KNP11}.

In analysis of MDPs, the behaviour under all possible strategies (schedulers, controllers, policies) is examined. 
For example, in the first two cases, the result of the verification process is either a guarantee that a given property holds under all strategies, or a counterexample strategy.
In the third case, either a witness strategy guaranteeing a given property is synthesized, or its non-existence is stated.
In all settings, it is desirable that the output \emph{strategies should be ``small and understandable''} apart from correct.
Intuitively, it is a strategy with a representation small enough for the human debugger to read and understand where the bug is (in the verification setting), or for the programmer to implement in the device (in the synthesis setting).
In this paper, we focus on the verification setting and illustrate our approach mainly on probabilistic protocols. 
Nonetheless, our results immediately carry over to the synthesis setting.

Obtaining a small and simple strategy may be impossible if the strategy is required to be optimal, i.e., in our setting reaching the error state with the highest possible probability. 
Therefore, there is a trade-off between simplicity and optimality of the strategies.
However, in order to debug a system, a simple counterexample or a series thereof is more valuable than the most comprehensive, but incomprehensible counterexample.
In practice, a simple strategy reaching the error with probability smaller by a factor of $\varepsilon$, e.g.\ one per cent, is a more valuable source of information than a huge description of an optimal strategy.
Similarly, controllers in embedded devices should be as optimal as possible, but only as long as they are small enough to fit in the device.
In summary, we are interested in finding small and simple close-to-optimal ($\varepsilon$-optimal) strategies.

How can one obtain a small and simple strategy? This seems to require some understanding of the particular system and the bug. How can we do something like that automatically? %Certainly, there is no algorithm to understand bugs. 
The approaches have so far been limited to BDD representation of the strategy, or generating subchains representing a subset of paths induced by the strategy. 
While BDDs provide a succinct representation, they are not well readable and understandable.
Further, subchains do not focus on the decisions the strategy makes at all.
In contrast, a~huge effort has been spent on methods to obtain ``understanding'' from large sets of data, using \emph{machine learning} methods. 
In this paper, we propose to extend their use in verification, namely of reachability properties, in several ways. 
Our first aim of using these methods is to efficiently exploit the structure that is present in the models, written in e.g.\ PRISM language with variables and commands. This structure gets lost in the traditional numerical analysis of the MDPs generated from the PRISM language description. 
The second aim is to distil more information from the generated MDPs, namely the importance of each decision.
% than just a probability of satisfying a given property.
Both lead to an improved understanding of the strategy's decisions.

\paragraph{\textbf{Our approach.}}
We propose three steps to obtain the desired strategies.
Each of them has a positive effect on the resulting size. %, and can be applied independently.
%However, only applying all three together yields strategies readable by human, according to our experimental results. 
We describe these three steps and illustrate them on an example.
% \begin{enumerate}
% \item \emph{Identifying important parts of the system.} \todo{leave out this item, later a sentence that applicable to extrememly large systems, again due to learning}
% As argued in \cite[our atva]{}, typically only a small fraction of the system needs to be explored in order to find an $\varepsilon$-optimal strategy.
% Intuitively, the reason is that most states are reached with only a very small probability, and whatever happens onwards does not affect the overall quality of the strategy too much. 
% In the MDP $M$ depicted in Fig.~\ref{fig:ex-intro}, a potentially huge part of the state space reachble from state $p$ is almost irrelevant for the overall probability of reaching $t$ from $s$.
% Consequently, the definition of the strategy on the unimportant parts of the state space may be safely ignored.

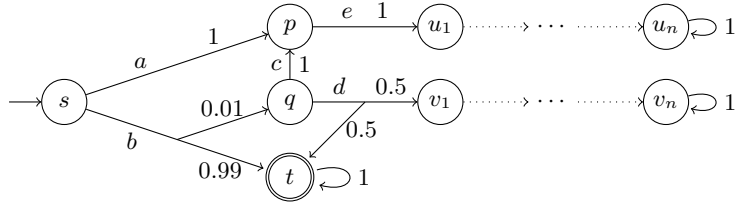
\begin{figure}[t]
\centering
\begin{tikzpicture}
\node[state,initial,initial text=] (s) at (-1,0){$s$};
\node[state] (r) at (2,1){$p$};
\node[state] (v1) at (4,1){$u_1$};
\node (vi) at (5.5,1){$\cdots$};
\node[state] (vn) at (7,1){$u_n$};
\node[coordinate] (c) at (0.5,-0.5) {};
\node[state] (xr) at (2,0){$q$};
\node[coordinate] (d) at(3,0) {};
\node[state] (xv1) at (4,0){$v_1$};
\node (xvi) at (5.5,0){$\cdots$};
\node[state] (xvn) at (7,0){$v_n$};
\node[state,accepting,outer sep=2] (t) at (2,-1){$t$};
\path[->] 
(s) edge node[above,pos=0.3]{$a$} node[above,pos=0.7]{$1$} (r)
(s) edge[-] node[below]{$b$} (c)
(c) edge node[above]{$0.01$} (xr)
(c) edge node[below]{$0.99$} (t)
(r) edge node[above]{$e\quad1$} (v1)
(v1) edge[dotted] node[above]{$ $} (vi)
(vi) edge[dotted] node[above]{$ $} (vn)
(vn) edge[loop right] node[right]{$1$} ()
(xr) edge[-] node[above]{$d$} (d)
(d) edge node[right]{$0.5$} (t)
(d) edge node[above]{$0.5$} (xv1)
(xr) edge node[left]{$c$} node[right]{$1$} (r)
(xv1) edge[dotted] node[above]{$ $} (xvi)
(xvi) edge[dotted] node[above]{$ $} (xvn)
(xvn) edge[loop right] node[right]{$1$} ()
(t) edge[loop right] node[right]{$1$} ()
;
\end{tikzpicture}
\caption{An MDP $M$ with reachability objective $t$}\label{fig:ex-intro}
\end{figure}

\smallskip\noindent\emph{(1)~Obtaining a (possibly partially defined and liberal) $\varepsilon$-optimal strategy.}
The $\varepsilon$-optimal strategies produced by standard methods, such as value iteration of PRISM, may be too large to compute and overly specific.
Firstly, as argued in \cite{atva}, typically only a small fraction of the system needs to be explored in order to find an $\varepsilon$-optimal strategy, whereas most states are reached with only a very small probability. Without much loss, the strategy may not be defined there.
For example, in the MDP $M$ depicted in Fig.~\ref{fig:ex-intro}, the decision in $q$ (and $v_i$'s) is almost irrelevant for the overall probability of reaching $t$ from $s$. Such a partially defined strategy can be obtained using learning methods~\cite{atva}.

Secondly, while the usual strategies prescribe which action to play, 
\emph{liberal} strategies leave more choices open. 
There are several advantages of liberal strategies, and similar 
notions of strategies called permissive strategies have been 
studied in~\cite{DBLP:journals/ita/BernetJW02,DBLP:conf/atva/BouyerMOU11,DBLP:conf/tacas/DragerFKPU14}. 
A liberal strategy, instead of choosing an action in each state, chooses a set of actions to be
played uniformly at every state.
First, each liberal strategy represents a set of strategies, and thus covers more behaviour. 
%Second, they allow more flexibility in the synthesis setting, since any strategy from the represented set can be implemented. 
Second, in counter-example guided abstraction-refinement (CEGAR) analysis, 
since liberal strategies can represent sets of counter-examples, they accelerate 
the abstraction-refinement loop by ruling out several counter-examples at once.
Finally, they also allow for more robust learning of smaller strategies in
Step 3.
We show that such strategies can be obtained from standard value iteration \cite{Kwiatkowska2013} as well as \cite{atva}. 
Further processing of the strategies in Step 2 and 3  allows liberal strategies as input and preserves liberty in the small representation of the strategy.

\smallskip\noindent\emph{(2)~Identifying important parts of the strategy.}
We define a concept of importance of a state w.r.t.\ a strategy, corresponding to the probability %/frequency 
of visiting the state by the strategy.
Observe that only a fraction of states can be reached while following the strategy, and thus have positive importance. On the unreachable states, with zero importance, the definition of the strategy is completely useless.
For instance, in $M$, both states $p$ and $q$ must have been explored when constructing the strategy in order to find out whether it is better to take action $a$ or $b$. 
However, if the resulting strategy is to use $b$ and $d$, the information what to do in $u_i$'s is useless.
In addition, we consider $v_i$'s to be of zero importance, too, since they are never reached on the way to target.

Furthermore, apart from ignoring states with zero importance, we want to partially ignore decisions that are unlikely to be made (in less important states such as $q$), and in contrast, stress more the decisions in important states likely to be visited (such as $s$). 
Note that this is difficult to achieve in data structures that remember all the stored data exactly, such as BDDs. Of course, we can store decisions in states with importance above a certain threshold. However, we obtain much smaller representation if we allow more variability and reflect the whole quantitative information, as shown in Step 3.

\smallskip\noindent\emph{(3)~Data structures for compact representation of strategies.} The explicit representation of a strategy by a table of pairs (state, action to play) results in a huge amount of data since the systems often have millions of states. 
Therefore, a symbolic representation by binary decision diagrams (BDD) looks as a reasonable option.
However, there are several drawbacks of using BDDs.
Firstly, due to the bit-level representation of the state-action pairs, the resulting BDD is not very readable.
Secondly, it is often still too large to be understood by human, for instance due to a bad ordering of the variables.
Thirdly, it cannot quantitatively reflect the differences in the importance of states.

Therefore, we propose to use \emph{decision trees} instead , e.g.~\cite{Mitchell1997}, a structure similar to BDDs, but with nodes labelled by various predicates over the system's variables. They have several advantages. 
Firstly, they yield an explanation of the decision, as opposed to e.g.\ neural networks, and thus provide an explanation how the strategy works.
Secondly, sophisticated algorithms for their construction, based on entropy, result in smaller representation than BDD, where a good ordering of variables is known to be notoriously difficult to find~\cite{BK08}.
Thirdly, as suggested in Step 2, they allow for less probable remembering of less stressed data if this sufficiently simplifies the tree and decreases its size.
Finally, the major drawback of decision trees in machine learning---frequent overfitting of the training data---is not an issue in our setting since the tree is not used for classification of test data, but only of the training data.
% \end{enumerate} 

\paragraph{Summary of our contribution.}
In summary our contributions are as follows:
\begin{compactitem}
\item We provide a method for obtaining succinct representation of $\varepsilon$-optimal strategies as decision trees. The method is based on a new concept of importance measure and on well-established machine learning techniques.
\item Experimental data show that even for some systems larger than the available memory, our method yields trees with only several dozens of nodes. 
\item We illustrate the understandability of the representation on several examples from PRISM benchmarks~\cite{KNP12b}, reading off respective bug explanations.
\end{compactitem}

\paragraph{Related work.}
%Compact representations of MDPs have been extensively studied in several contexts. 
In artificial intelligence, compact (factored) representations of MDP structure have been developed using dynamic Bayesian networks~\cite{BDG95,KHW94}, probabilistic STRIPS~\cite{KK99}, algebraic decision diagrams~\cite{HSHB99}, and also decision trees~\cite{BDG95}. 
Formalisms used to represent MDPs can, in principle, be used to represent values and policies as well. In particular, variants of decision trees are probably the most used~\cite{BDG95,CK91,KP99}.
% (there are other possible representations such as predictive policy representations~\cite{}). 
For a detailed survey of compact representations see~\cite{BDH99}.
In the context of verification, MDPs are often represented using variants of (MT)BDDs~\cite{dAKNPS00,HKNPS03,MP04}, and strategies by BDDs~\cite{WBB+10}.

%Decision trees have been used in connection with real-time dynamic programming (reinforcement learning) to represent the learned approximation of the value function~\cite{BD96,DBLP:conf/appinf/Pyeatt03}. Learning a compact decision tree representation of a policy has been investigated in~\cite{LPRT10} for the case of body sensor networks. This paper is closest to our work but aims at a completely different application field (a simple model of sensor networks as opposed to generic PRISM models), uses different objectives (discounted rewards as opposed to unbounded reachability), and does not consider priorities that, as we show, may substantially decrease sizes of resulting policies.

Decision trees have been used in connection with real-time dynamic programming and reinforcement learning~\cite{BD96,DBLP:conf/appinf/Pyeatt03}. Learning a compact decision tree representation of a policy has been investigated in~\cite{LPRT10} for the case of body sensor networks, but the~paper aims at a completely different application field (a simple model of sensor networks as opposed to generic PRISM models), uses different objectives (discounted rewards),
% as opposed to unbounded reachability), 
and does not consider the importance of a state that, as we show, may substantially decrease sizes of resulting policies.

%
% This paper is the closest to our work but aims at a completely different application field (a simple model of sensor networks as opposed to generic PRISM models), uses different objectives (discounted rewards as opposed to unbounded reachability), and does not consider the importance of a state that, as we show, may substantially decrease sizes of resulting policies.

Our results are related to the problem of computing minimal counterexamples in probabilistic verification. Most papers concentrate on solving this problem for Markov chains and linear-time properties~\cite{HKD09,ADR08,WJAKB14,JAKWKB11}, branching-time properties~\cite{DHKK08,FHPW10,AL10}, and in the context of simulation~\cite{KPC12}.
%For branching-time properties,~\cite{} proposes to compute counterexamples for PCTL formulae via games. 
A~couple of tools have been developed for probabilistic counterexample generation, namely DiPro~\cite{ALLS11} and COMICS~\cite{JAVWKB12}.
 For a detailed survey see~\cite{ABDJKW14}.

%The problem of computing minimal counterexamples has been studied for various stochastic systems and specification formalisms. Most papers concentrate on solving this problem for Markov chains. As mentioned in~\cite{WJAKB14}, for linear-time properties several approaches have been investigated, such as computation of (an~appropriate representation of) sufficiently many (un)successful paths~\cite{HKD09,ADR08} and computation of so called minimal critical subsystems, i.e., minimal sub-chains disproving a given property~\cite{WJAKB14,JAKWKB11}. Small counter-example generation has also been investigated for branching-time properties, expressed mostly in PCTL~\cite{DHKK08,FHPW10,AL10}, and in the context of simulation~\cite{KPC12}.
%%For branching-time properties,~\cite{} proposes to compute counterexamples for PCTL formulae via games. 
%A couple of tools have been developed for probabilistic counterexample generation, namely DiPro~\cite{ALLS11} and COMICS~\cite{JAVWKB12}.
% For a detailed survey see~\cite{ABDJKW14}.

Concerning MDPs,~\cite{WJAKB14} uses mixed integer linear programming to compute minimal critical sub-systems, i.e. whole sub-MDPs as opposed to a compact representation of %a~sequence 
``right" decisions computed by our methods. 
\cite{AL09} uses a directed on-the-fly search to compute sets of most probable diagnostic paths (which somehow resembles our notion of importance), but the paths are encoded explicitly by AND/OR trees as opposed to our use of decision trees. 
Neither of these papers takes advantage of an~internal structure of states and their methods substantially differ from ours.
The notion of paths encoded as AND/OR trees has also been studied in~\cite{LL13} to represent probabilistic counter-examples visually as fault trees, and then derive causal (the cause and effect) relationship between events.
\cite{KH09} develops abstraction-based framework for model-checking MDPs based on games, which allows to trade compactness for precision, but does not give a procedure for constructing (a compact representation of) counterexample strategies. 
% and their methods are based on completely different principles than methods of this paper. 
\cite{DJWAK14} computes a smallest set of guarded commands (of a~PRISM-like language) that induce a critical subsystem, but, unlike our methods, does not
provide a compact representation of actual decisions needed to reach an erroneous state and is incomplete as there is not always a command based counterexample. 
%To the~best of our knowledge, none of these works uses machine learning methods to obtain a compact representation of counterexamples.
Moreover, all previous works considered a single strategy, and none of them considered computation and representation of liberal strategies.

Counter-examples play a crucial role in CEGAR analysis of MDPs, and have been
widely studied, such as, game-based abstraction refinement~\cite{KNP06};
non-compositional CEGAR approach  for reachability~\cite{HWZ08} and safe-pCTL~\cite{CV10};
compositional CEGAR approach for safe-pCTL and qualitative logics~\cite{KPC12,CCD14}; and 
abstraction-refinement for quantitative properties~\cite{DArgenioJJL01,DArgenioJJL02}. 
All of these works only consider a single strategy represented explicitly, whereas our 
approach considers a succinct representation of a set of strategies, and can accelerate the 
abstraction-refinement loop.

%!TEX root = main.tex

\section{Preliminaries}

We use  $\Nset$, $\Qset$, and $\Rset$ to denote the sets of positive integers, rational and real numbers, respectively. 
The set of all rational probability distributions over a finite set $X$ is denoted by $\dist(X)$.
Further, $d\in\dist(X)$ is Dirac if $d(x)=1$ for some $x\in X$. Given a function $f:X\rightarrow \Rset$, we  
% denote by $\argmax_{x\in X} f(x)$ an $x\in X$ satisfying $f(x)=\max_{x'\in X} f(x')$.
write $\argmax_{x\in X} f(x)=\{x\in X\mid f(x)=\max_{x'\in X} f(x')\}$.

\smallskip\noindent{\bf Markov chains.} 
A \emph{Markov chain} is a tuple \mbox{$M = (L,P,\mu)$} where $L$ is a finite %countable 
set of locations, 
$P:L\to\dist(L)$ is a probabilistic transition function, and
$\mu\in\dist(L)$ is the initial probability distribution.

A \emph{run} in $M$ is an infinite sequence $\pat = \ell_1 \ell_2 \cdots$ of locations,
a \emph{path} in $M$ is a finite prefix of a run. 
Each path $\fpat$ in $M$ determines the set $\Cone(\fpat)$ consisting of all runs that start with $\fpat$. 
To $M$ we associate the probability space $(\Pat,\calF,\pr)$, 
where $\Pat$ is the set of all runs in $M$, $\calF$ is the $\sigma$-field generated by all $\Cone(\fpat)$,
and $\pr$ is the unique probability measure such that
$\pr[\Cone(\ell_1\cdots\ell_k)] = 
\mu(\ell_1) \cdot \prod_{i=1}^{k-1} P(\ell_i)(\ell_{i+1})$.

%For $G\subseteq S$, we denote the set of runs that reach $G$ by $\reach G:=\{\ell_1 \ell_2 \cdots\in\Pat\mid\exists t:\ell_t\in G\}$.

\smallskip\noindent{\bf Markov decision processes.} 
A \emph{Markov decision process} (MDP) is a tuple $G=(S,A,\act,\delta,\inits)$ 
where $S$ is a finite set of states, $A$ is a finite set of actions, 
$\act : S\rightarrow 2^A\setminus \{\emptyset\}$ assigns to each state $s$ the set $\act(s)$ of actions enabled 
in $s$, % so that $\{\act{s}\mid s\in S\}$ is a partitioning of $A$,
$\delta : S\times A\rightarrow \dist(S)$ is a probabilistic 
transition function that, given a state and 
an action, 
%$a \in \act{s}$ enabled at $s$ 
gives a probability distribution over the 
successor states, and $\inits$ is the initial state.
%Note that we consider that every action is enabled in exactly one state.

A \emph{run} in $G$ is an infinite alternating sequence of states
and actions $\pat=s_1 a_1 s_2 a_2\cdots$
such that for all $i \geq 1$, we have $a_i\in\act(s_i)$ and $\delta(s_i,a_i)(s_{i+1}) > 0$. 
%We denote by $\Pat_G$ the set of all runs in~$G$.
A \emph{path} of length~$k$ in~$G$ is a finite prefix
$\fpat = s_1 a_1\cdots a_{k-1} s_k$ of a run in~$G$.
%For a finite path $\fpat$ we denote by $\last(\fpat)$ the last state of~$w$.

\smallskip\noindent{\bf Strategies and plays.} 
Intuitively, a strategy (or a policy) in an MDP $G$ is a ``recipe'' to choose actions.
Formally, a strategy is a function $\sigma:S\to \dist(A)$  that given the current state of a play
gives a probability distribution over the enabled actions.\footnote{%
In general, a strategy is  a function $\sigma : (SA)^*S \to \dist(A)$ that given a finite path~$\fpat$, representing 
the history of a play, gives a probability distribution over the 
actions enabled in the last state. However, for objectives considered in this paper, these more general strategies are not more powerful than our restricted \emph{memoryless} strategies (depending on the last state visited).
In order to simplify the notation, we thus only consider memoryless strategies in this paper.}
In general, a strategy may randomize, i.e.\ return non-Dirac distributions. 
A strategy is \emph{deterministic} if it gives a Dirac distribution for every argument.

%A \emph{play} of $G$ determined by 
%a strategy $\sigma$ is a Markov chain 
%$G^\sigma$ where the set of locations is $S$,
%the initial distribution $\mu$ is Dirac with $\mu(\inits)=1$ and
%$$P(s)(s')=\sum_{a\in A}\sigma(s)(a) \cdot \delta(s,a)(s')\,.$$
%The induced probability measure is denoted by $\pr^{\sigma}$ and ``almost surely'' or ``almost all runs'' refers to happening with probability 1 according to this measure.
%%The respective expected value of a random variable $f:\Pat\to\Rset$ is $\Ex \sigma f=\int_\Pat f\ d\,\pr^\sigma$.
%%For $t\in\Nset$, the random variable $S_t$ returns the $t$-th location on the run. 

A \emph{play} of $G$ determined by 
a strategy $\sigma$ and a state $\bar{s}\in S$ is a Markov chain 
$G^\sigma_{\bar{s}}$ where the set of locations is $S$,
the initial distribution $\mu$ is Dirac with $\mu(\bar{s})=1$ and
$$P(s)(s')=\sum_{a\in A}\sigma(s)(a) \cdot \delta(s,a)(s')\,.$$
The induced probability measure is denoted by $\pr^{\sigma}_{\bar{s}}$ and ``almost surely'' or ``almost all runs'' refers to happening with probability 1 according to this measure. We usually write $\pr^{\sigma}$ instead of $\pr^{\sigma}_{\inits}$ (here $\inits$ is the initial state of $G$).
%The respective expected value of a random variable $f:\Pat\to\Rset$ is $\Ex \sigma f=\int_\Pat f\ d\,\pr^\sigma$.
%For $t\in\Nset$, the random variable $S_t$ returns the $t$-th location on the run. 
%

%LIBERAL
%{\color{blue}
%A {\em liberal strategy} is a function $\lstrat:S\rightarrow 2^A$ such that for every $s\in S$ we have that $\lstrat(s)\in 2^{\act(s)}\smallsetminus \{\emptyset\}$. A strategy $\sigma$ is an {\em instance} of the liberal strategy $\lstrat$ if for every $s\in S$ we have $\lstrat(s)=\{a\in \act(s)\mid \sigma(s)(a)>0\}$. A liberal strategy $\lstrat$ is $\varepsilon$-optimal if every instance of $\lstrat$ is $\varepsilon$-optimal. Given a liberal strategy $\lstrat$ and a state $s$, an action $a\in \act(s)$ is {\em good} (in $s$ w.r.t. $\lstrat$) if $(s,a)\in \lstrat(s)$, and {\em bad} otherwise. We denote by $\sigma^{u}_{\lstrat}$ the strategy which always plays all good actions uniformly, i.e., for every $s\in S$ we have that $\sigma^{u}_{\lstrat}(s)$ is the~uniform distribution on $\varsigma(s)$.
%}
%{\color{blue}

\smallskip\noindent\textbf{Liberal strategies.}
A {\em liberal strategy} is a function $\lstrat:S\rightarrow 2^A$ such that for every $s\in S$ we have that $\emptyset\not = \lstrat(s)\subseteq \act(s)$. 
%$\lstrat(s)\in 2^{\act(s)}\smallsetminus \{\emptyset\}$. 
Given a liberal strategy $\lstrat$ and a state $s$, an action $a\in \act(s)$ is {\em good} (in $s$ w.r.t. $\lstrat$) if $a\in \lstrat(s)$, and {\em bad} otherwise.
Abusing notation, we denote by $\lstrat$ the strategy that to every state $s$ assigns the uniform distribution on $\lstrat(s)$ 
(which, in particular, allows us to use $G^{\varsigma}_s$, $\pr^{\varsigma}_s$ and apply the notion of $\varepsilon$-optimality to $\varsigma$).
%
%Abusing notation, we denote by $\lstrat$ the strategy which plays all good actions uniformly (which, in particular, allows us to use $G^{\varsigma}_s$, $\pr^{\varsigma}_s$ and apply the notion of $\varepsilon$-optimality to $\varsigma$).
%We say that $\lstrat$ is $\varepsilon$-optimal if $\sigma_{\lstrat}$ is $\varepsilon$-optimal.
%}
%\todo{It is possible to generalize a bit and say that $\varsigma$ denotes an arbitrary but fixed strategy, which plays exactly good actions with positive probabilities(?)}
%/LIBERAL

\smallskip\noindent{\bf Reachability objectives.}
Given a set $F\subseteq S$ of {\em target states}, we denote by $\reach{F}$ the set of all runs %initiated in $\inits$ 
that visit a state of $F$.
For a state $s\in S$, the \emph{maximal reachability probability} (or simply {\em value}) in $s$, is 
$\val(s):=\max_{\sigma} \pr^{\sigma}_s[\reach{F}]$. 
%\todo{$\pr^{\sigma}[\reach{T}\mid\reach s]$}\todotomas{Just added the state index to $\pr^{\sigma}_{\bar{s}}$ for MDPs, in my opinion, it is not a big complication. The index may be omitted if $\bar{s}=\inits$.}
%
Given $\epsilon\geq 0$, we say that a strategy $\sigma$ is {\em $\varepsilon$-optimal} if $\pr^{\sigma}[\reach{F}] \geq \val(\inits)-\varepsilon$,
and we call a $0$-optimal strategy {\em optimal}.\footnote{For every MDP, there is a memoryless deterministic optimal strategy, see e.g.~\cite{Puterman:book}.}
To avoid overly technical notation, we assume that states of $F$, subject to the reachability objective, are absorbing, i.e.\ for all $s\in F,a\in\act(s)$ holds $\delta(s,a)(s)=1$.

\smallskip\noindent{\bf End components.} 
A non-empty set $S'\subseteq S$
is an \emph{end component} (EC) of $G$ if there is $\mathit{Act}':S'\to 2^A\setminus\{\emptyset\}$ such that 
(1) for all $s\in S'$ we have $\mathit{Act}'(s)\subseteq\act(s)$,
(2) for all $s\in S'$, we have $a\in \mathit{Act}'(s)$ iff $\delta(s,a)\in\dist(S')$, %$\delta(s,a)(s')>0$ implies $s'\in S'$,
and (3) for all $s,t\in S'$ there is a path 
$\pat = s_1 a_1\cdots a_{k-1} s_k$ such that $s_1 = s$, $s_k=t$, and $s_i\in S',a_i\in\mathit{Act}'(s_i)$ for every $i$. 
An end component is a \emph{maximal end component} (MEC)
if it is maximal with respect to the subset ordering. Given an MDP, the set of MECs is denoted by $\mec$.
Given a MEC, actions of $\mathit{Act}'(s)$ and $\act(s)\setminus\mathit{Act}'(s)$ are called \emph{internal} and \emph{external} (in $s$), respectively.
\section{Computing $\varepsilon$-optimal Strategies}
%There are many algorithms for solving quantitative reachability in Markov decision processes, such as the value iteration, the strategy improvement, linear programming based methods etc. (For details on classical solution methods for MDPs see~\cite{Puterman:book}.)
%%
% The main method implemented in PRISM is the value iteration which successively (under)approximates the~value $\val(s,a)=\sum_{s'\in A} \delta(s,a)(s')\cdot \val(s')$ of every state-action pair $(s,a)$ by a value $V(s,a)$, and stops when the approximation is good enough. More precisely, denoting by $V(s):=\max_{a\in \act(s)} V(s,a)$, every step of the value iteration improves the approximation $V(s,a)$ by assigning $V(s,a):=\sum_{s'\in S} \delta(s,a)(s')\cdot V(s')$ (the computation starts with $V$ such that $V(s)=1$ if $s\in F$, and $V(s)=0$ otherwise).
 %
There are many algorithms for solving quantitative reachability in Markov decision processes, such as the value iteration, the strategy improvement, linear programming based methods etc., see~\cite{Puterman:book}.
 The main method implemented in PRISM is the value iteration, which successively (under)approximates the~value $\val(s,a)=\sum_{s'\in A} \delta(s,a)(s')\cdot \val(s')$ of every state-action pair $(s,a)$ by a value $V(s,a)$, and stops when the approximation is good enough. Denoting by $V(s):=\max_{a\in \act(s)} V(s,a)$, every step of the value iteration {\em improves} the approximation $V(s,a)$ by assigning $V(s,a):=\sum_{s'\in S} \delta(s,a)(s')\cdot V(s')$ (we start with $V$ such that $V(s)=1$ if $s\in F$, and $V(s)=0$ otherwise).

%The disadvantage of the standard value iteration (and also most of the above mentioned traditional methods) is that it works with the whole state space of the MDP (or at least with its reachable part). 
%For instance, consider states $u_i,v_i$ of Fig.~\ref{fig:ex-intro}. 
%The paper~\cite{atva} proposes to adapt methods of bounded real-time dynamic programming (BRTDP, see~e.g.~\cite{}) to speed up the computation of the value iteration by improving values of $V(s,a)$ only on ``important'' parts of the state space. In every iteration, the~important parts of the state space are identified using simulated executions of a strategy  induced by the current approximation of values $V$ (roughly speaking, the strategy chooses actions with maximum value of the current approximation). The paper~\cite{atva} shows that this method may substantially reduce the number of state-action pairs for which the value needs to be stored, which also results in a smaller representation of the~resulting strategy (see~Section~\ref{sec:exper} for comparison of the size of the reachable state space and the~space explored by the simulations). This method also gives an approximation $V(s,a)$ of the value $\val(s,a)$ for every state-action pair $(s,a)$ (more precisely, using the notation of~\cite{atva}, as $V$ we use the lower approximation $V_L$).

The disadvantage of the standard value iteration (and also most of the above mentioned traditional methods) is that it works with the whole state space of the MDP (or at least with its reachable part). 
For instance, consider states $u_i,v_i$ of Fig.~\ref{fig:ex-intro}. 
The paper~\cite{atva} adapts methods of bounded real-time dynamic programming (BRTDP, see~e.g.~\cite{MBLG05}) to speed up the computation of the value iteration by improving $V(s,a)$~\footnote{Here we use $V$ for the lower approximation denoted by $V_L$ in~\cite{atva}.} only on ``important'' state-action pairs identified by simulations.
% of strategies determined by the current approximation of $V$.
% In every iteration, the~important parts of the state space are identified using simulated executions of a strategy  that chooses actions with maximum value $V$.

%{\color{blue}
Even though RTDP methods may substantially reduce the size of an $\varepsilon$-optimal strategy, its explicit representation is usually large and difficult to understand. Thus we develop succinct representations of strategies, based on decision trees, that will reduce the size even further and also provide a~human readable representation. 
Even though the above methods are capable of yielding {\em deterministic} $\varepsilon$-optimal strategies, that can be immediately fed into machine learning algorithms, we found advantageous to give the learning algorithm more freedom in the sense that if there are more $\varepsilon$-optimal strategies, we let the algorithm choose (uniformly). This is especially useful within end-components where many actions have the same value. 
Therefore, we extract {\em liberal} $\varepsilon$-optimal strategies from the value approximation $V$, output either by the value iteration or BRTDP.
%}
%
%As most machine learning methods can be directly used for function approximation, they may, in principle, be directly applied to any $\varepsilon$-optimal strategy. However, we found advantageous to give the learning algorithm more freedom in the sense that if there are more $\varepsilon$-optimal strategies, we let the algorithm choose. This is especially useful within end-components where many actions have the same value. 
%{\color{blue} 
%Therefore, we extract {\em liberal} $\varepsilon$-optimal strategies from the value approximation $V$, output either by the value iteration or BRTDP.
%}
%Therefore, we extract the following representation of $\varepsilon$-optimal strategies from the value approximation $V$, output either by the value iteration or BRTDP.

\subsubsection{Computing liberal $\varepsilon$-optimal strategies.}
%Given $s\in S$ and $a\in A$, we denote by $Q(s,a)\in \{\g,\bad\}$ the {\em quality} of the action $a$ in $s$. We say that a~strategy $\sigma$ {\em conforms to $Q$} if for every $s\in S$ and $a\in A$ holds $\sigma(s)(a)>0$ iff $Q(s,a)=\g$.\footnote{This definition may be relaxed a bit by assuming that (1) $\sigma(s)(a)>0$ implies $Q(s,a)=\g$, and (2) if $s$ is in a MEC and $Q(s,a)=\g$, then $\sigma(s)(a)>0$. The reason for (2) is that we need to make sure that all states of a MEC are visited almost surely.}
% 
%%Intuitively, if $Q(s,a)=\g$, then taking $a$ in $s$ is possible in any $\varepsilon$-optimal strategy. 
%We say that $Q$ is {\em $\varepsilon$-optimal} if any $\sigma$ that conforms with $Q$ is $\varepsilon$-optimal.

Let us show how to obtain a liberal strategy $\lstrat$ from the value iteration, or BRTDP. For simplicity, we start with MDP without MECs.

\paragraph{MDP without end components.}  Say that $V:S\times A\rightarrow [0,1]$ is a {\em valid $\varepsilon$-underapproximation} if the following conditions hold:
\begin{compactenum}
\item $V(s,a)\leq \val(s,a)$ for all $s\in S$ and $a\in A$
\item $\val(\inits)-V(\inits)\leq \varepsilon$
\item $V(s,a)\leq \sum_{s'\in S} \delta(s,a)(s')\cdot V(s')$ for all $s\in S$ and $a\in \act{s}$
\end{compactenum}
The outputs $V$ of both the value iteration, and BRTDP are valid $\varepsilon$-underapproximations.
We define a liberal strategy $\lstrat^V$ by 
$\lstrat^V(s)=\argmax_{a\in \act(s)}\, V(s,a)$ for all $s\in S$.
\begin{lemma}\label{lem:approx-opt}
For every $\varepsilon>0$ and a valid $\varepsilon$-underapproximation $V$,
 $\lstrat^V$ is $\varepsilon$-optimal.~\footnote{ 
Intuitively this means that randomizing among good actions of $\varepsilon$-optimal strategies preserves $\varepsilon$-optimality in the reachability setting 
(in contrast to other settings, e.g.\ with parity objectives).}
%Let $V$ be the resulting approximation of $\val$ out of the value iteration. Then any strategy $\sigma_V$ that for every state $s$ chooses an action out of $\argmax_{a\in \act{s}} V(s,a)$ is $\varepsilon$-optimal.
\end{lemma}
\paragraph{General MDP.} For MDPs with end components we have to extend the definition of the~valid $\varepsilon$-underapproximation. Given a~MEC $S'\subseteq S$, we say that 
 $(s,a)\in S\times A$ is {\em maximal-external in $S'$} if $s\in S'$, $a\in \act(s)$ is external
  and $V(s,a)\geq V(s',a')$ for all $s'\in S'$ and $a'\in \act(s')$. A state $s'\in S'$ 
  is an~{\em exit} (of $S'$) if $(s,a)$ is ext-max in $S'$ for some $a\in \act(s)$.
We add the following condition to the valid $\varepsilon$-underapproximation:
\begin{compactenum}
\item[4.] Each MEC $S'\subseteq S$ has at least one exit.
% $s\in S'$ and an external $a\in\act(s)$ such that for all $s'\in S'$ and all internal $a'\in \act(s')$ we have $V(s,a)\geq V(s',a')$
\end{compactenum}
Now the definition of $\lstrat^V$ is also more complicated:
\begin{compactitem}
\item For every $s\in S$ which is {\em not} in any MEC, we put $\lstrat^V(s)=\argmax_{a\in \act(s)}\, V(s,a)$.
\item For every $s\in S$ which {\em is} in a MEC $S'$, 
	\begin{compactitem}
	\item if $s$ is exit, then $\lstrat^V(s)=\{a\in \act(s)\mid (s,a)\text{ is maximal-external in }S'\}$
	\item otherwise, $\lstrat^V(s)=\{a\in \act(s)\mid a\text{ is internal}\}$
	\end{compactitem}
\end{compactitem}
Using these extended definitions, Lemma~\ref{lem:approx-opt} remains valid.
Further, note that $\lstrat^V(s)$ is defined even for states with trivial underapproximation $V(s)=0$, for instance a state $s$ that was never subject to any value iteration improvement. Then the values $\lstrat(s)$ may not be stored explicitly, but follow implicitly from \emph{not} storing any $V(s)$, thus assuming $V(s,\cdot)=0$.

\section{Importance of Decisions}\label{ssec:importance-of-decisions}
%Assume that we have computed an $\varepsilon$-optimal liberal strategy $\lstrat$.
%
%Now we may, in principle, compute a succinct representation of $\lstrat$, or of any instance of $\lstrat$ (using e.g. BDDs), and obtain a strategy with possibly smaller representation than above. 

Note that once we have computed an $\varepsilon$-optimal liberal strategy $\lstrat$, we may, in principle, compute a compact representation of $\lstrat$ (using e.g. BDDs), and obtain a strategy with possibly smaller representation than above. 

However, we go one step further as follows. Given a liberal strategy $\lstrat$ and a state $s\in S$, we define the \emph{importance} of $s$ by 
$$\imp^\lstrat(s):=\mathbb{P}^{\lstrat}[\reach s\mid \reach F]$$ 
the probability of visiting $s$ conditioned on reaching $F$ (afterwards). Intuitively, the importance is high for states where a good decision can help to reach the target.\footnote{Instead of the conditional probability of reaching $s$, we could consider the conditional expected number of visits of $s$. We discuss the differences and compare the efficiency together with the case of no conditioning on reaching the target in Section~\ref{sec:exper}.}

\begin{example}
For the MDP of Fig.~\ref{fig:ex-intro} with the objective $\reach \{t\}$ and strategy $\lstrat$ choosing $b$, we have $\imp^{\lstrat}(s)=1$ and $\imp^{\lstrat}(q)=5/995$. Trivially, $\imp^\lstrat(t)=1$. For all other states, the importance is zero.
\end{example}
Obviously, decisions made in states of zero importance do not affect 
$\pr^{\lstrat}[\reach{F}]$ since these states never occur on paths from $\inits$ to $F$.
However, note that many states of $S$ may be reachable in $G^{\lstrat}$ with positive but negligibly small probability.
Clearly, the value of $\pr^{\lstrat}[\reach{F}]$ depends only marginally on choices made in these states. 
Formally, 
% given $\delta>0$ and a strategy $\lstrat$, we say that a state $s$ is {\em $(\delta, \lstrat)$-important} if the~probability of visiting $s$ before $F$ is at least $\delta$.
let $\lstrat_{\thresh}$ be a strategy obtained from $\lstrat$ by changing each $\lstrat(s)$ with $\imp^\lstrat(s)\leq \thresh$ to an arbitrary subset of $\act(s)$.
%
%Then every strategy $\lstrat$ satisfies 
%$\displaystyle\lim_{\thresh\rightarrow 0} \pr^{\lstrat_{\thresh}}[\reach{F}]=\pr^{\lstrat}[\reach{F}]$
%the probability $\pr^{\lstrat_{\thresh}}[\reach{F}]$ monotonically approaches $\pr^{\lstrat}[\reach{F}]$ (from below) as $\thresh$ goes to $0$.
%we have $\displaystyle\lim_{\delta\rightarrow 0} \pr^{\lstrat_{\delta}}[\reach{F}]=\pr^{\lstrat}[\reach{F}]$. 
%\end{lemma}
%Of course, for every strategy $\lstrat$, 
%In fact, every $\thresh<\min_{s\in S} \imp^\lstrat(s)$ satisfies $\pr^{\lstrat_{\thresh}}[\reach{F}]=\pr^{\lstrat}[\reach{F}]$. But often even larger $\thresh$ may give $\pr^{\lstrat_{\thresh}}[\reach{F}]$ sufficiently close to $\pr^{\lstrat}[\reach{F}]$. Such $\thresh$ may be found using e.g.\ trial and error approach.\footnote{One may even give a (quite conservative) bound on convergence of $\pr^{\lstrat_{\thresh}}[\reach{F}]$ to $\pr^{\lstrat}[\reach{F}]$ as $\thresh\rightarrow 0$, using e.g. Lemma~5.1 of~\cite{BKK14}. However, for large MDPs the bound would be impractical.}
%
We obtain the following obvious property: %It is obvious that 

\begin{lemma}
For every liberal strategy $\lstrat$, 
%the probability $\pr^{\lstrat_{\thresh}}[\reach{F}]$ monotonically approaches $\pr^{\lstrat}[\reach{F}]$ (from below) as $\thresh$ goes to $0$.
we have $\displaystyle\lim_{\thresh\rightarrow 0} \pr^{\lstrat_{\thresh}}[\reach{F}]=\pr^{\lstrat}[\reach{F}]$. 
\end{lemma}
In fact, every $\thresh<\min(\{\imp^\lstrat(s)\mid s\in S\}\setminus\{0\})$ satisfies $\pr^{\lstrat_{\thresh}}[\reach{F}]=\pr^{\lstrat}[\reach{F}]$. But often even larger $\thresh$ may give $\pr^{\lstrat_{\thresh}}[\reach{F}]$ sufficiently close to $\pr^{\lstrat}[\reach{F}]$. Such $\thresh$ may be found using e.g.\ trial and error approach.\footnote{One may even give a (quite conservative) bound on convergence of $\pr^{\lstrat_{\thresh}}[\reach{F}]$ to $\pr^{\lstrat}[\reach{F}]$ as $\thresh\rightarrow 0$, using e.g. Lemma~5.1 of~\cite{BKK14}. However, for large MDPs the bound would be impractical.}

%\begin{lemma}
%For every strategy $\sigma$ we have $\displaystyle\lim_{\delta\rightarrow 0} \pr^{\sigma_{\delta}}[\reach{F}]=\pr^{\sigma}[\reach{F}]$. 
%\end{lemma}
%Therefore, if $\sigma$ is $\varepsilon$-optimal, then for every $\varepsilon'>\varepsilon$ there is $\delta>0$ such that $\sigma_{\delta}$ is $\varepsilon'$-optimal. In principle, it is possible to give (quite conservative) bounds on $\delta$ using standard theory of Markov chains~\todotomas{Do this at least in app?}. Another option is to use trial and error approach to find the optimal $\delta$ that cuts off as many states as possible, but does not affect the precision of the strategy too much.

Most importantly,
%Furthermore, 
we can use the importance of a state to affect the probability that decisions in this state are indeed remembered in the data structure.
Data structures with such a feature are used in various learning algorithms. In the next section, we discuss decision trees.
Due to this extra variability, which decisions to learn, the resulting decision trees are smaller than BDDs for strictly defined $\lstrat_\thresh$.

\section{Efficient Representations}\label{ssec:efficient-representations}

Let $G=(S,A,\mathit{Act},\delta,\inits)$ be an MDP. In order to symbolically represent strategies in $G$, we need to assume that states and actions have some internal structure. Inspired by PRISM language \cite{KNP11}, we consider a set $\mathcal{V}=\{v_1,\ldots,v_n\}$ of {\em integer variables}, each $v_i$ gets its values from a finite domain $\mathit{Dom}(v_i)$. We suppose that $S=\prod_{i=1}^n \mathit{Dom}(v_i)\subseteq \Zset^n$. Further, we assume that the MDP arises as a product of $m$ modules, each of which can separately perform non-synchronizing actions as well as synchronously with other modules perform a synchronizing action. Therefore, we suppose $A\subseteq\bar A \times \{0,\ldots,m\}$, where $\bar A\subseteq \Nset$ is a finite set and the second component determines the module performing the action ($0$ stands for synchronizing actions).\footnote{On the one hand, PRISM does not allow different modules to have local variables with the same name, hence we do not distinguish which module does a variable belong to. On the other hand, while PRISM declares no names for non-synchronizing actions, we want to exploit the connection between the corresponding actions of different copies of the same module.}
% \todotomas{Explain relationship with the PRISM code: For variables, we ignore which module it comes from, i.e., each $v_i$ corresponds either to a global variable $x$ in PRISM, or to $(x,m)$ where $x$ is a local variable of the module $m$. For actions, we {\em do not} ignore the module (i.e., $(a,m)\in A\subseteq \Nset\times \Nset$ where $a$ is the action in PRISM code, $m$ is the module. For synchronizing actions we assume a default "module".}

%Since a liberal strategy is a function of the form $\sigma:S \rightarrow 2^A$, assigning to each state its good actions,
%it can be {\em explicitly} represented as a list of state-action pairs, i.e., as a subset of 
%{\color{blue}
Since a liberal strategy is a function of the form $\lstrat:S \rightarrow 2^A$, assigning to each state its good actions,
it can be {\em explicitly} represented as a list of state-action pairs, i.e., as a subset of 
%}
\begin{equation}\label{eq:domain}
S\times A=\prod_{i=1}^n \mathit{Dom}(v_i)\times \bar A \times \{0,1,\ldots,m\}
\end{equation}
In addition, standard optimization algorithms implemented in PRISM use an explicit ``don't-care'' value $-2$ for action in each unreachable state, meaning the strategy is not defined. 
However, one could simply not list these pairs at all. Thus a smaller list is obtained, with only the states where $\varsigma$ is defined.
%Note that such a representation can be applied to $\sigma_0$, thus ignoring also reachable states with zero probability to reach the target.
%{\color{blue} 
Recall that one may also omit states $s$ satisfying $\imp^{\lstrat}(s)=0$, thus ignoring  reachable states with zero probability to reach the target.
Further optimization may be achieved by omitting states $s$ satisfying $\imp^{\lstrat}(s)<\Delta$ for a suitable $\Delta>0$.
%}

%{\color{blue} Note that such a representation can be applied to $\lstrat_\Delta$ with $\Delta=0$, thus ignoring also reachable states with zero probability to reach the target.}
%Furthermore, it can be applied to $\lstrat_\thresh$ for $\thresh>0$, resulting in a shorter list. 
%In Section~\ref{sec:exper}, we compare the savings achieved on several examples.
%\todo{no experiments done}

\subsection{BDD Representation}

The explicit set representation can be encoded as a binary decision diagram (BDD). This has been used in e.g.~\cite{WBB+10,EJPV12}.
The principle of the BDD representation of a set is that (1) each element is encoded as a string of bits and (2) an automaton, in the form of a binary directed acyclic graph, is created so that (3) the accepted language is exactly the set of the given bit strings.
Although BDDs are quite efficient, see Section~\ref{sec:exper}, each of these three steps can be significantly improved: 
\begin{compactenum}
\item Instead of a string of bits describing all variables, a string of integers (one per variable) can be used. Branching is then done not on the value of each bit, but according to an inequality comparing the variable to a constant. This significantly improves the readability. 

\item Instead of building the automaton according to a chosen order of bits, we let a heuristic choose the order of the inequalities and the actual constants in the inequalities. 

\item Instead of representing the language precisely, we allow the heuristic to choose which data to represent and which not. The likelihood that each datum is represented corresponds to its importance, which we provide as another input.
\end{compactenum}
The latter two steps lead to significantly smaller graphs than BDDs.
All this can be done in an efficient way using decision trees learning.

\subsection{Representation using Decision Trees}

\subsubsection{Decision trees.}

A \emph{decision tree} for a domain $\prod_{i=1}^d X_i\subseteq \Zset^d$ is a tuple $\mathcal{T}=(T,\rho,\theta)$ where $T$ is a finite rooted binary (ordered) tree with a set of inner nodes $N$ and a set of leaves $L$, $\rho$ assigns to every inner node a predicate of the form $[x_i\sim \mathit{const}]$ where $i\in\{1,\ldots,d\}$, $x_i\in X_i$, $\mathit{const}\in \Zset$, $\mathord{\sim}\in\{\leq,<,\geq,>,=\}$, and $\theta$ assigns to every leaf a value $\true$, or $\false$.~\footnote{There exist many variants of decision trees in the literature allowing arbitrary branching, arbitrary values in the leaves, etc., e.g.~\cite{Mitchell1997}.
For simplicity, we define only a special suitable subclass.}

Similarly to BDDs, the language $\mathcal{L}(\mathcal{T})\subseteq \Nset^n$ of the tree is defined as follows. For a vector $\vec{\bar x}=(\bar x_1,\ldots,\bar x_n)\in \Nset^n$, we find a path $p$ from the root to a leaf such that for each inner node $n$ on the path, the predicate $\rho(n)$ is satisfied by substitution $x_i=\bar x_i$ iff the~first child of $n$ is on $p$. Denote the leaf on this particular path by $\ell$. Then $\vec{\bar x}$ is in the language $\mathcal{L}(\mathcal{T})$ of $\mathcal T$ iff $\theta(\ell)=\true$.

\begin{example}
Consider dimension $d=1$, domain $X_1=\{1,\ldots,7\}$. A tree representing a set $\{1,2,3,7\}$ is depicted in Fig.~\ref{fig:dectree}. To depict the ordered tree clearly, we use unbroken lines for the first child, corresponding to the satisfied predicate, and dashed line for the second one, corresponding to the unsatisfied predicate.  
\begin{figure}
\centering
\begin{tikzpicture}[node distance = 1.2cm]
	\node (a1) [draw,rectangle,rounded corners=2pt] {$x_1\leq 3$} ;
	\node (a2) [below right of=a1,draw,rectangle,rounded corners=2pt]  {$x_1<7$};
	\node (good1) [below right of=a2,draw,rectangle,rounded corners=2pt] {$\true$};
	\node (bad1) [below left of=a2,draw,rectangle,rounded corners=2pt] {$\false$};
	\node (good2) [below left of=a1,draw,rectangle,rounded corners=2pt] {$\true$};
	
	\draw [-,dashed] (a1) to (a2);
	\draw [-] (a1) to (good2);
	\draw [-,dashed] (a2) to (good1);
	\draw [-] (a2) to (bad1);
\end{tikzpicture}
	\caption{A decision tree for $\{1,2,3,7\}\subseteq\{1,\ldots,7\}$}
	\label{fig:dectree}
\myspace	
\end{figure}
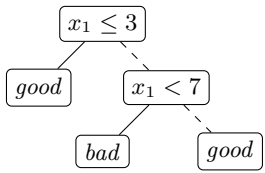
\end{example}

In our setting, we use the domain $S\times A$ defined by Equation~(\ref{eq:domain}) which is of the form $\prod_{i=1}^{n+2} X_i$ where for each $1\leq i\leq n$ we have $X_i=\mathit{Dom}(v_i)$, $X_{n+1}=\bar{A}$ and $X_{n+2}=\{0,1,\ldots,m\}$.
Here the coordinates $\mathit{Dom}(v_i)$ are considered ``unbounded'' and, consequently, the respective predicates use inequalities.
In contrast, we know the possible values of $\bar A\times \{0,1,\ldots,m\}$ in advance and they are not too many. 
Therefore, these coordinates are considered ``discrete'' and the respective predicates use equality. 
Examples of such trees are given in Section~\ref{sec:exper} in Fig.~\ref{fig:zeroconftree_1} and~\ref{fig:mertree}. 
%{\color{blue}
Now a decision tree $\mathcal{T}$ for this domain determines a liberal strategy $\lstrat:S\rightarrow 2^A$ by $a\in \lstrat(s)$ iff $(s,a)\in \mathcal{L}(\mathcal{T})$.
%}
%For the sake of readability, instead of the equalities, we use the keyword $\play$.
%For the sake of readability, instead of equalities for actions, we use the keyword $\play$ (an action).

\subsubsection{Learning.}

We describe the process of \emph{learning a training set}, which can also be understood as storing the input data.
Given a training sequence (repetitions allowed!) $\vec x^1,\ldots,\vec x^k$, with each $\vec x^i=(x^i_1,\ldots,x^i_n)\in \Nset^d$, partitioned into the \emph{positive} and \emph{negative} subsequence, the process of learning according to the algorithm ID3 \cite{Quin86,Mitchell1997} proceeds as follows: 	 
	 	\begin{compactenum}
	 	\item Start with a single node (root), and assign to it the whole training sequence.
	 	\item Given a node $n$ with a sequence $\tau$,
	 	\begin{compactitem}
%	 	\item if it is purely positive, set $\theta(n)=\true$ and stop;
%	 	\item if it is purely negative, set $\theta(n)=\false$ and stop;
	 	\item if all training examples in $\tau$ are positive, set $\theta(n)=\true$ and stop;
	 	\item if all training examples in $\tau$ are negative, set $\theta(n)=\false$ and stop;
	 	\item otherwise, 
			\begin{compactitem} 	
		 	\item choose a predicate with the ``highest gain'' (with lowest entropy, see e.g.~\cite[Sections 3.4.1, 3.7.2]{Mitchell1997}), 
		 	\item split $\tau$ into sequences satisfying and not satisfying the predicate, assign them to the first and the second child, respectively, 
		 	\item go to step 2 for each child.
		 \end{compactitem}
	 	\end{compactitem}
	 	\end{compactenum}
Intuitively, the predicate with the highest gain splits the sequence so that it maximizes the portion of positive data in the satisfying subsequence and the portion of negative data in the non-satisfying subsequence.

In addition, the final tree can be \emph{pruned}. This means that some leaves are merged, resulting in a smaller tree at the cost of some imprecision of storing (the language of the tree changes). 
The pruning phase is quite sophisticated, hence for the sake of simplicity and brevity, we omit the details here. 
We use the standard C4.5 algorithm and refer to~\cite{Quinlan93,Mitchell1997}. 
In Section~\ref{sec:exper}, we comment on effects of parameters used in pruning.

\subsubsection{Learning a strategy.}
Assume that we already have a liberal strategy $\lstrat:S\rightarrow 2^A$. 
%We say that a state-action pair $(s,a)$ is {\em good} if $a\in \lstrat(s)$ and {\em bad} otherwise.
We show how we learn good and bad state-action pairs so that the language of the resulting tree is close to the set of good pairs.
The training sequence will be composed of state-action pairs where good pairs are
positive examples, and bad pairs a negative ones.
Since our aim is to ensure that important states are learnt and not pruned away, we repeat pairs with more important states in the training sequence more frequently. 

Formally, for every $s\in S$ and $a\in \act(s)$, we %want to
 put the pair $(s,a)$ to the training sequence $\mathit{repeat}(s)$-times, where
$$\mathit{repeat}(s)=c\cdot\imp^\lstrat(s)$$
for some constant $c\in\Nset$ (note that $\imp^\lstrat(s)\leq 1$).
Since we want to avoid exact computation of $\imp^\lstrat(s)$, we estimate it using simulations.
%Since we do not know the value $\imp^\sigma(s)$, we estimate it using simulations. 
In practice, we thus run $c$ simulation runs that reach the target and set $\mathit{repeat}(s)$ to be the number of runs where $s$ was also reached.

\section{Experiments}\label{sec:exper}

%We discuss the examples and show that the resulting decision trees can be read manually to help what part of the model allows it to reach the target.
%
%We compare three different data-structures for representing strategies, which are a table representation, BDDs and decision trees.
%On of our goals is to produce human readable trees and therefore want to minimize the size of the tree.
%Therefore, we present the respective sizes of the data structures for each example.
In this section, we present the experimental evaluation of the presented methods, which we have implemented within the probabilistic 
model checker PRISM \cite{KNP11}.  All the results presented in this section were obtained on a single Intel(R) Xeon(R) CPU (3.50GHz) with memory limited to 10GB.

First, we discuss several alternative options to construct the training data and to learn them in a decision tree. %s from the Weka package given an MDP.
Further, we compare decision trees to other data structures, namely sets and BDDs, with respect to the sizes necessary for storing a strategy.
Finally, we illustrate how the decision trees can be used to gain insight into our benchmark examples.

\subsection{Generating Training Data}
The strategies we work with come from two different sources.
%We use two sources of strategies that are used later as an input for decision tree construction.
% We use two kinds of strategies as input for decision tree construction.
Firstly, we consider strategies constructed by PRISM, which can be generated using the explicit or sparse model checking engine.
%, or the sparse engine under restricted action label conditions, which are not satisfied in three of our four examples we consider.\todojan{what? is restricted whatever important? if it is not satisfied %it means we're doing something incorrect?}
Secondly, we consider strategies constructed by the BRTDP algorithm~\cite{atva}, which are defined on a part of the state space only. 
%that are given implicitly by the heuristic computation.
%The use of explicit model checking is only feasible for smaller examples, as illustrated in Table~\ref{tab:sizeresults}.

Recall that given a strategy, the training data for the decision trees is constructed from~$c$ simulation runs according to the strategy.
In our experiments, we found that $c=10000$ produces good results in all the examples we consider.
Note that we stop each simulation as soon as the target or a state with no path to the target state is reached.
% During the simulation the algorithm keeps track of the number of visits to a state and what are the good and bad actions for every state reached by the simulation.
% This data is used later for different learning variants.
% The alternative approach is to construct the training data by iterating over the whole state space rather than doing simulations.
% However, this approach did not yield promising results.

\subsection{Decision Tree Learning in Weka}

The decision trees are constructed using the Weka machine learning package \cite{Hall2009}.
The Weka suite offers various decision tree classifiers.
We use the J48 classifier, which is an implementation of the C4.5 algorithm \cite{Quinlan93}.
The J48 classifier offers two parameters  to control the pruning that affect the size of the decision tree:
%$M$ is a positive integer value that specifies the minimum number of instances in a leaf of the decision tree.
\begin{compactitem}
\item Firstly, the leaf size parameter $M$ determines that each leaf node with less than $M$ instances in the training data is merged with its siblings.
The value $M$ can be any positive integer. However, only values smaller than the minimum number of instances per
 classification class are reasonable, since higher numbers always result in the trivial tree of size $1$.
\item The confidence factor $C$ is used internally for determining the amount of pruning during decision tree construction.
The value of $C$ can be any double value in the half open interval $(0,0.5]$. Smaller values incur more pruning and therefore smaller trees.%\todojan{really? smaller values are more precise, aren't they?}
\end{compactitem}
More information and an empirical study of the parameters for J48 can be found in~\cite{Drazin2012}.

\smallskip\noindent\textbf{Effects of the parameters.}
We illustrate the effects of the parameters $C$ and $M$ on the resulting size of the decision tree on the \texttt{mer} benchmark. 
However, similar behaviour appears in all the examples.
Figures~\ref{fig:merconf} and~\ref{fig:mermin} show the resulting size of the decision tree. % for the \texttt{mer} example.
Each line in the plots corresponds to one decision tree, learned with $15$ different values of the parameter.
The $C$ parameter scales linearly between $0.0001$ and $0.5$.
The $M$ parameter scales logarithmically between $1$ and the minimum number of instances per class in the respective training set.
The plots in Figure~\ref{fig:params} show that $M$ is an effective parameter in calibrating the resulting tree size, whereas $C$ plays less of a role. Hence, we use $C=10^{-4}$.
Furthermore, since the tree size is monotone in $M$, the parameter $M$ can be used to retrieve a desired level of detail from tree.%\todojan{fill in M,C for rel. error}

\myspace
\begin{center}
\begin{figure}[h!t]
	\begin{subfigure}[t]{0.3\textwidth}
		\centering
		\includegraphics[scale=0.13]{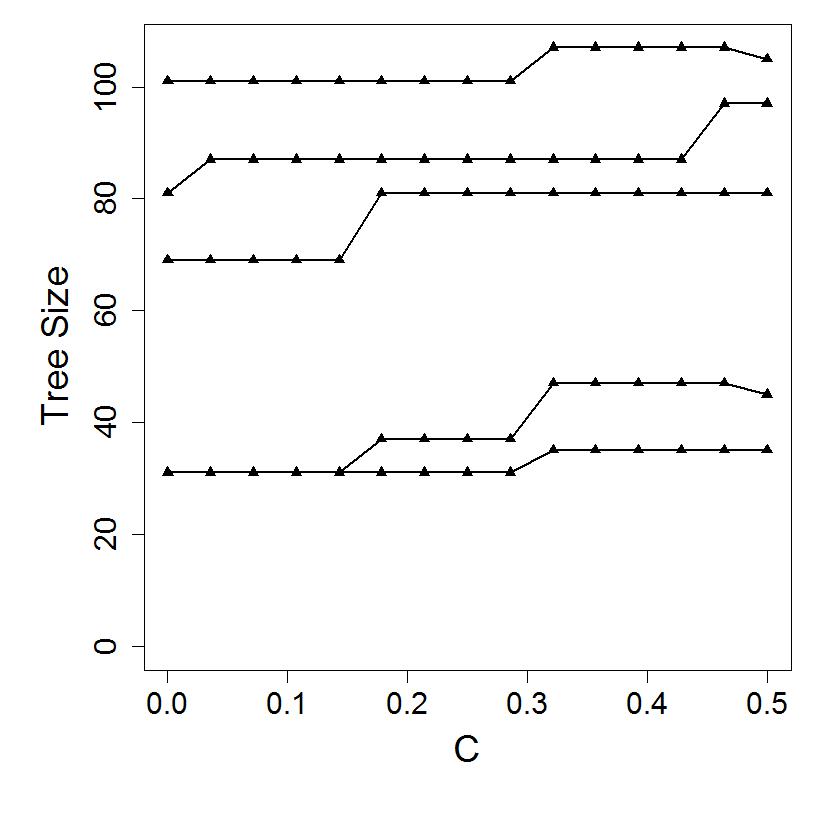}
		\caption{fixed $M = 2$}
		\label{fig:merconf}
	\end{subfigure}
	\begin{subfigure}[t]{0.3\textwidth}
		\includegraphics[scale=0.13]{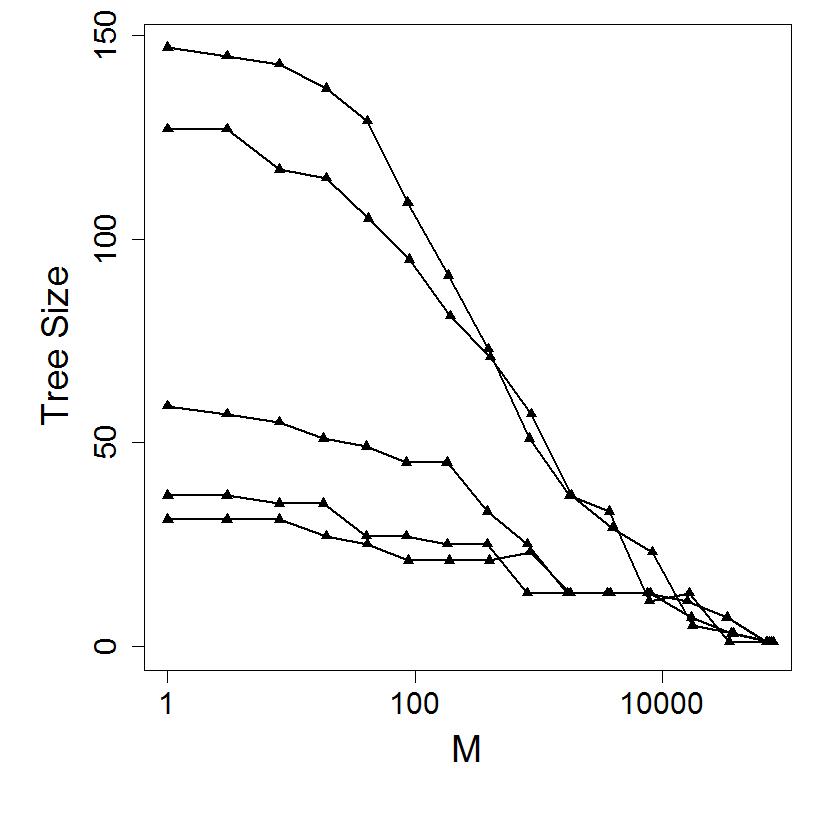}
		\caption{fixed $C = 10^{-4}$}
		\label{fig:mermin}
	\end{subfigure}
	\begin{subfigure}[t]{0.3\textwidth}
		\includegraphics[scale=0.13]{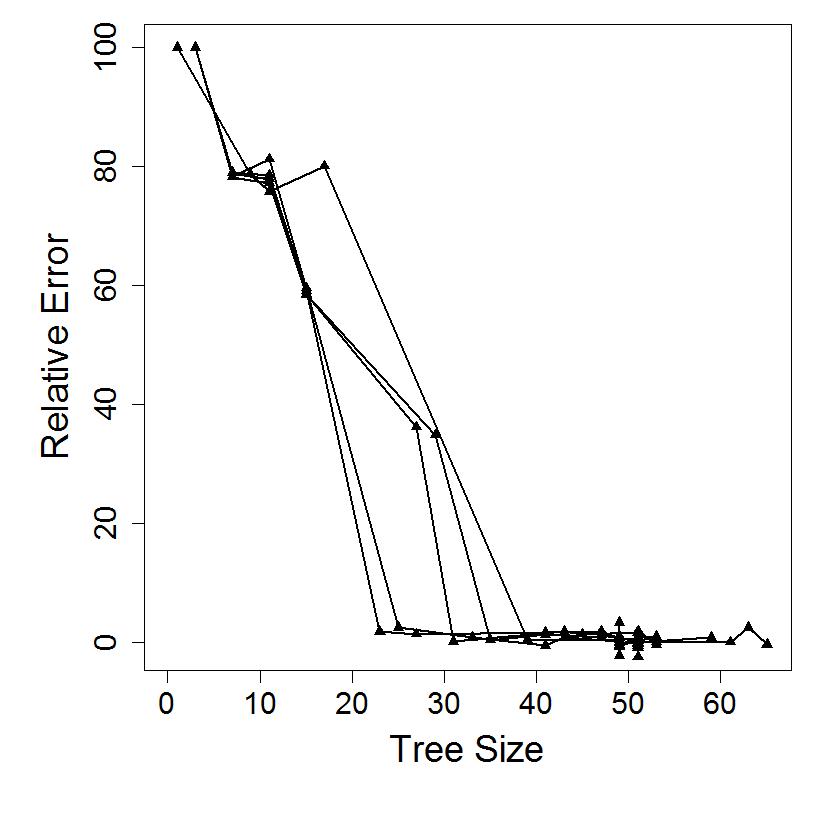}
		\caption{Tree Size vs Error}
		\label{fig:mererror}
	\end{subfigure}	
	\caption{Decision tree Parameters}
	\label{fig:params}
\end{figure}
\myspace\myspace\myspace\myspace
\end{center}

Figure~\ref{fig:mererror} depicts the relation of the tree size to the relative error of the induced strategy. 
%The error is computed from the maximum reachability probability of the strategy and the results obtained by the BRTDP computation. 
% For smaller examples, the maximum reachability probability can be obtained by transforming the decision tree into a classic table represented strategy and model check the Markov chain that is the product of this strategy and the original model.
% For larger examples, the maximum reachability probability can be approximated by running simulations that count how often the target state is reached.
%Figure~\ref{fig:mererror} 
It shows that there is a threshold size under which the tree is not able to capture the strategy correctly anymore and the error rises quickly.
Above the threshold size, the error is around $1\%$, %which we consider a good characterization for the strategy 
considered reasonable in order to extract reliable information.
This threshold behaviour is observed in all our examples.
% and together with the effectiveness of the parameter $M$ in pruning trees, motivates 
Therefore, it is sensible to perform a binary search for the highest $M$ ensuring the error at most $1\%$ and we do so in the next section. 
% that produces a tree with size over this threshold.

\subsection{Results}

\setlength{\tabcolsep}{3pt}

\begin{savenotes}
\begin{table}[t]
\caption{Comparison of representation sizes for strategies obtained from PRISM and from BRTDP computation. Sizes are presented as the number of states for explicit lists of values, the number of nodes for BDDs, and the number of nodes for decision trees (DT). For DT, we display the tree size obtained from the binary search described above. Error reports the relative error of the strategy determined by the decision tree compared to the optimal value, obtained by model checking with PRISM.}
\label{tab:sizeresults}

\centering
\begin{threeparttable}
	\resizebox{\linewidth}{!}{
	\begin{tabular}{lrl|rrrl|rrrl}
	\toprule
	  & & &  \multicolumn{4}{c|}{\textbf{PRISM}} & \multicolumn{4}{c}{\textbf{BRTDP}} \\
		\textbf{Example} & {$\vert S \vert$}  & Value & Explicit & BDD & DT & Error & Explicit & BDD & DT & ~Error \\ 
		\midrule
		firewire & 481,136 & 1.000 &479,834 & 4,233 & 1 & 0.000\%& 766 & 4,763 \tnote{1} & 1 & ~0.000\% \\
		investor & 35,893 & 0.958 & 28,151 & 783 & 27 & 0.886\% & 21,931& 2,780 & 35 & ~0.836\%\\
		mer\_17M & 1,773,664 & 0.200 & \multicolumn{4}{c|}{Memory Out} & 1,887 & 619 & 17 & ~0.000\% \\
		mer\_big~\tnote{2} & Approx. $10^{13}$ & 0.200 & \multicolumn{4}{c|}{Memory Out} & 1,894 & 646 & 19 & ~0.692\% \\
		zeroconf & 89,586 & 0.009 & 60,463 &  409 & 7 & 0.106\% & 1,630& 905 & 7 & ~0.235\% \\
	\bottomrule
	\end{tabular}
	}
\begin{tablenotes}
\item[1] Note that BDDs represent states in binary form. Therefore, one entry in the explicit state list corresponds to several nodes in the BDD.
\item[2] We did not measure the state size as the MDP does not fit in memory, but extrapolated it from the linear dependence of model size and one of its parameters, which we could increase to $2^{31}-1$. The value is obtained from the BRTDP computation.
\end{tablenotes}
\end{threeparttable}
\end{table}
\end{savenotes}
\setlength{\tabcolsep}{0pt}

First, we briefly introduce the four examples from the PRISM benchmark suite \cite{KNP12b}, which we tested our method on:
%These examples come with PRISM.

% \begin{compactitem}
\smallskip\noindent\textbf{firewire}
models the Tree Identify Protocol of the IEEE 1394 High Performance Serial Bus, which is used to transport video and audio signals within a network of multimedia devices.
The reachability objective is that one node gets the root and the other one the child role.

\smallskip\noindent\textbf{investor}
models a market investor and shares, which change their value probabilistically over time.
The reachability objective is that the investor finishes a trade at a time, when his shares are more valuable than some threshold value.

\smallskip\noindent\textbf{mer}
is a mutual exclusion protocol, that regulates the access of two users to two different resources.
The protocol should prohibit that both resources are accessed simultaneously.
\begin{wraptable}[11]{r}{5.5cm}
 	\caption{Effects of various learning variants on the tree size. Smallest trees computed from PRISM or BRTDP are presented.}
	\label{tab:learning}
	\centering
	\begin{tabular}{c|cccccc}	
	Example & $\textsc{I}\reach\pr$ & $\textsc{I}\forall\pr$ & $\textsc{I}\reach\expected$ & $\textsc{I}\forall\expected$ & $\textsc{O}\reach$ & $\textsc{O}\forall$\\
	\toprule
 	firewire & 1 & 1 & 1 & 1 & 1 & 1 \\
 	investor &  27 & 25 & 31 & 35 & 37 & 37 \\
 	mer\_17M & 17 & 33 & 17 & 29 & 19 & none \\
	mer\_big & 19 & 23 & 23 & 37 & 17 & none \\
 	zeroconf   & 7 & 7 & 7 & 7 & 7 & 17 \\	
 	\bottomrule
 	\end{tabular}

\end{wraptable}

\smallskip\noindent\textbf{zeroconf}
is a network protocol which allows users to choose their IP addresses autonomously.
% Conflicts are detected via broadcast messages.
The protocol should detected  and prohibit IP address conflict.
% \end{compactitem}

For every example, Table~\ref{tab:sizeresults} shows the size of the state space, the value of the optimal strategy, and
 the sizes of strategies obtained from explicit model checking by PRISM and by BRTDP, for each discussed data structure. 
 %\todojan{what about the trillion states example? do we comment anywhere? no graph needed, just a sentence}
 %\todojan{table says mer 13, the figure ater has 17!!! If 13 is not sure, we have to write 17 even in the table}
 %\todojan{brtdp firewire - how can BDD be much larger than the explicit list?!}

\smallskip\noindent\textbf{Learning variants.}
% In Section \ref{ssec:importance-of-decisions} the importance of a state was defined.
% In the experiments, we reflect this measure by constructing the training data in different ways.
In order to justify our choice of the importance function $\imp$, we compare it to several alternatives.
%There exist different approaches to construct the training data from the simulations. In our work we consider three boolean options that result in a total number of six possible combinations:
\begin{compactenum}
\item When constructing the training data, we can use the importance measure $\imp$, and add states as often as is indicated by its importance ($\textsc{I}$), or neglect it and simply add every visited state exactly once ($\textsc{O}$).
\item Further, states on the simulation are learned conditioned on the fact that the target state is reached ($\reach$). Another option is to consider all simulations ($\forall$).
\item Finally, instead of the probability to visit the state ($\pr$), one can consider the expected number of visits ($\expected$).
%\item The importance measure $\imp$ for weighting states can be used ($\textsc{I}$) or it can be neglected so every state that was counted at least once is added to the training set exactly once.
\end{compactenum}
In Table \ref{tab:learning}, we report the sizes of the decision trees obtained for the all learning variants. We conclude that our choice ($\textsc{I}\reach\pr$) is the most useful one.
%\todojan{can you reproduce 13 or not? if not then put bigger number! what is the star? you really can't get precise enough tree?}

\iffalse %%%%%%%%%%% the $\thresh$ is a DIFFERENT $\delta$ !!!
\smallskip\noindent\textbf{Suboptimal actions by varying parameter $\thresh$.}
The strategy obtained from the heuristic computation of \cite{atva} is represented by lower and upper bounds on the reachability probability of actions.
These values can be used to learn not only the best actions, but also consider suboptimal ones with their lower bound in the $\thresh$ range of the best one.
This approach allows us to capture a set of strategies at once.
In Table~\ref{tab:delta}, we show the effects of increasing the $\thresh$ parameter both in terms of ratio of good actions per state and the resulting size of the tree.

\begin{table}
	\centering
	\begin{tabular}{ccc|cc|cc|cc}
	\toprule
	 & \multicolumn{2}{c}{\textbf{firewire}}& \multicolumn{2}{c}{\textbf{investor}} & \multicolumn{2}{c}{\textbf{mer}} & \multicolumn{2}{c}{\textbf{zeroconf}} \\
	$\thresh$ & good actions & DT	& good actions & DT& good actions & DT& good actions & DT\\ 
	\midrule
	0 & & & 65\% & 129 & 73\% & 25 & 70\% & 7  \\
	0.01 & & & 77\% & 41 & 80\% & 51 & 83\% & 7 \\
	0.1 & & & 78\% & 137 & 79\% & 43 & 86\% & 7 \\
 	0.5 & & & 83\% & * & 98\% & * & 88\% & * \\
 	\bottomrule
 	\end{tabular}
 	\vspace{0.2em}
	\caption{Effects of $\thresh$ \\ (* no dec. tree with less than 1\% error.)}
	\label{tab:delta}
\end{table}
\fi

\subsection{Understanding Decision Trees}

We show how the constructed decision trees can help us to gain insight into the essential features of the systems.

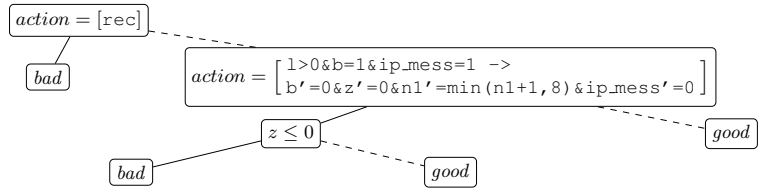
\begin{figure}
	\centering
	%!TEX root = ../main.tex

% \documentclass{standalone}
% \usepackage{tikz}
%\def\checkmark{\tikz\fill[scale=0.4](0,.35) -- (.25,0) -- (1,.7) -- (.25,.15) -- cycle;} 

% \begin{document}

%played_action = rec
%|   b_ip0 <= 0.0: bad (21.0/3.0)
%|   b_ip0 > 0.0: good (2.0)
%played_action != rec
%|   played_action = [] l>0&b=0&n0>0 -> 0.6 : (b'=2) & (ip_mess'=0) & (n0'=n0-1) + 0.4 : (n0'=n0-1): bad (11.0/3.0)
%|   played_action != [] l>0&b=0&n0>0 -> 0.6 : (b'=2) & (ip_mess'=0) & (n0'=n0-1) + 0.4 : (n0'=n0-1): good (127.0/22.0)
%
\resizebox{0.8\linewidth}{!}{
\begin{tikzpicture}[node distance = .9cm]
	\node (a1) [draw,rectangle,rounded corners=2pt] {\playaction{rec}};

	\node (a2) [below right of=a1,draw,rectangle,rounded corners=2pt,yshift=-.3cm,align=left,xshift=5.5cm] {$action = \Big[$ \begin{tabular}{l} \texttt{l>0\&b=1\&ip\_mess=1 ->} \\ \texttt{b'=0\&z'=0\&n1'=min(n1+1,8)\&ip\_mess'=0} \end{tabular} $\Big]$};
	\node (bad1) [left =of a2,draw,rectangle,rounded corners=2pt,xshift=-1cm] {$\false$};
	\node (good1) [below right of=a2,draw,rectangle,rounded corners=2pt,yshift=-.3cm,xshift=4cm] {$\true$};
	
	\node (a3) [below left of=a2,draw,rectangle,rounded corners=2pt,yshift=-.3cm,xshift=-2cm] {$z \leq 0$};
	\node (bad2) [below left of=a3,draw,rectangle,rounded corners=2pt,xshift=-2cm] {$\false$};
	\node (good2) [below right of=a3,draw,rectangle,rounded corners=2pt,xshift=2cm] {$\true$};
	
	\draw [-] (a1) to (bad1);
	\draw [-,dashed] (a1) to (a2);
	
	\draw [-] (a2) to (a3);
	\draw [-,dashed] (a2) to (good1);

	\draw [-] (a3) to (bad2);	
	\draw [-,dashed] (a3) to (good2);

\end{tikzpicture}
}

% \end{document}
	\caption{A decision tree for {\texttt zeroconf}}
	\label{fig:zeroconftree_1}
	\myspace\myspace
\end{figure}

\smallskip\noindent\textbf{\textbf{\texttt{zeroconf} example.}}
In Figure \ref{fig:zeroconftree_1} we present a decision tree that is a strategy for \texttt{zeroconf} and shows how an unresolved IP address conflict can occur in the protocol.
First we present how to the read the strategy represented in Figure~\ref{fig:zeroconftree_1}. Next we show how the strategy can explain the conflict in the protocol.
Assume that we are classifying a state-action pair $(s,a)$, where action $a$ is enabled in state $s$.
\begin{compactenum}

\item No matter what the current state $s$ is, the action \texttt{rec} is always classified as $\false$ according to the root of the tree. Therefore, the action \texttt{rec} should be played with positive
probability only if all other available actions in the current state are also classified as $\false$.

\item If action $a$  is different from \texttt{rec}, the right son of the root node is reached.
 If action $a$ is different from action {\tt l>0\&b=1\&ip\_mess=1 -> b'=0\&z'=0\&n1'=min(n1+1,8)\&ip\_mess'=0} (the whole PRISM command is a single action),
 then $a$ is classified as $\true$ in state $s$. Otherwise, the left son is reached.

\item In node ${\tt z \leq 0}$ the classification of action $a$ (that is the action that labels the parent node) depends on the variable valuation of the current state. If the value of var. $z$ is greater than $0$, then
 $a$ is classified as $\true$ in state $s$, otherwise it is classified as $\false$.
 % is enabled in all states that satisfy the precondition {\tt l > 0 \& b=1 \& ip\_mess=1}. If the variable valuation in the current state allows this ac
\end{compactenum}
% and should only be played if no other $\true$ action is available in the current state.

Action \texttt{rec} stands for a network host receiving a reply to a broadcast message, resulting in resolution of an IP address conflict if one is present, which clearly does not help in constructing an unresolved conflict.
The action labelling the right son of the root  represents the detection of an IP address conflict by an arbitrary network host.
This action is only good, if variable $z$, which is a clock variable, in the current state is greater than $0$.
The combined meaning of the two nodes is that an unresolved IP address conflict can occur if the conflict is detected too late.

%The decision trees for this example can be trimmed down to the trivial tree and still remain exact precision.
%This indicates that the system is robust in the sense that no wrong decision can be made and the objective will be reached no matter which strategy is played.
\smallskip\noindent\textbf{\textbf{\texttt{firewire} example.}} For {\tt firewire}, we obtain a trivial tree with a single node, labelled $\true$.
Therefore, playing all available actions in each state guarantees reaching the target almost surely.
%This is a strong statement about the robustness of the protocol. 
In contrast to other representations, we have automatically obtained the~information that the network always reaches the target configuration, regardless of the individual behaviour of its components and their interleaving.

\smallskip\noindent\textbf{\textbf{\texttt{mer} example.}}
In the case of \texttt{mer}, there exists a strategy that violates the required property that the two resources
are not accessed simultaneously. 
%In the case of the PRISM model checker the computed strategy does not fit in the memory. In the
%case of BRTDP, the computed strategy represented either explicitly or with BDDs, is too large to be understandable. 
The decision tree for the \texttt{mer} strategy is depicted in Figure~\ref{fig:mertree}. 
In order to understand how a state is reached, where both resources are accessed at the same time, it is necessary 
to determine which user accesses which resource in that state.
% This information is contained in nodes marked with $1$ and $2$ respectively.

\begin{compactenum}
\item The two tree nodes labelled by $1$ explain what resource \emph{user 1} should access. The root node labelled by action 
{\tt s1=0\&r1=0 -> r1'=2} specifies
 that the request to access \emph{resource $2$}  (variable {\tt r1} is set to $2$) is classified as $\false$. The only remaining action
 for \emph{user 1} is to request access to \emph{resource $1$}. This action is classified as $\true$ by the right son of the root node. 

\item Analogously, the tree nodes labelled by $2$ specify that \emph{user 2} actions should request access to \emph{resource 2}
(follows from {\tt s2=0\&r2=0 -> r2'=2}). Once \emph{resource 2} is requested it should change its internal state {\tt s2}
to $1$  (follows from {\tt s2=0\&r2=2 -> s2'=1}).
 It follows,
that in the state violating the property, \emph{user~1} has access to \emph{resource 1} and \emph{user 2} to \emph{resource 2}.

\end{compactenum}

The model is supposed to correctly handle such overlapping requests, but fails to do so in a specific case.
In order to further debug the model, one has to find the action of the scheduler that causes this undesired behaviour.
The lower part of the tree specifies that {\tt u1\_request\_comm} is a candidate for such an action.
Inspecting a snippet of the  code of {\tt u1\_request\_comm} from the PRISM source code (shown below) reveals that in the given
 situation, the scheduler reacts inappropriately with some probability $p$.
%\todojan{the first ... stands for a guard. what about the other two?}
\begin{verbatim}	
	[u1_request_comm]  s=0 & commUser=0 & driveUser!=0 & k<n ->  
	           (1-p):(s'=1) & (r'=driveUser) & (k'=k+1) + 
	               p:(s'=-1) & (gc'=true) & (k'=k+1)
\end{verbatim}

	% {\tt[u1 request comm]  s=0 & commUser=0 & driveUser <= 0  & k < n ->  
	% 				   & (1-p):(s'=1) & (r'=driveUser) & (k'=k+1) + 
	% 				   & p:(s'=-1) & (gc'=true) & (k'=k+1)}

%Indeed this command allows to reach an undesired state with probability $p$.
%The line also explains the two remaining internal nodes of the decision tree.
%In case the $(1-p)$ part of the command was executed, the situation should be reset and the action in the fourth internal node is necessary to do so.
%The erroneous line can only be executed $n$ times counted up by $k$.
%Therefore, for big enough $k$, every action leads equally (un)likely to the target state.
The remaining nodes of the tree that were not discussed are necessary to reset the situation if the non-faulty part (with probability $1-p$) 
 of the {\tt u1\_request\_comm} command was executed.
It should be noted that executing the faulty {\tt u1\_request\_comm} action does not lead to the undesired state right away.
The action only grants \emph{user 1} access rights in a situation, where he should not get these rights.
Only a successive action leads to \emph{user 1} accessing the resource and the undesired state being reached.
This is a common type of bug, where the command that triggered an error is not the cause of it.

\begin{center}
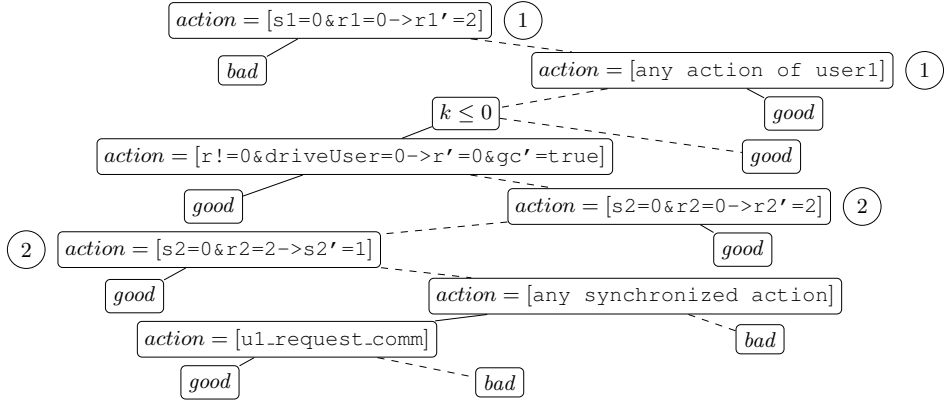
\begin{figure}
	%!TEX root = ../main.tex

%\documentclass{standalone}
%\usepackage{tikz}
%\def\checkmark{\tikz\fill[scale=0.4](0,.35) -- (.25,0) -- (1,.7) -- (.25,.15) -- cycle;} 
%\newcommand{\true}{\mathsf{tt}}
%\newcommand{\false}{\mathsf{ff}}

%\begin{document}
\resizebox{\linewidth}{!}{
\begin{tikzpicture}[node distance = .9cm]
	\node (a1) [draw,rectangle,rounded corners=2pt] {\playaction{s1=0\&r1=0->r1'=2}};

	\node (a2) [below right of=a1,draw,rectangle,rounded corners=2pt,xshift=5cm,yshift=-0.1cm] {\playaction{any action of user1}};
	\node (bad1) [left =of a2,draw,rectangle,rounded corners=2pt,xshift=-3cm]  {$\false$};

	\node (a3) [below left of=a2,draw,rectangle,rounded corners=2pt,xshift=-3cm]  {$k \leq 0$};
	\node (good1) [right =of a3,draw,rectangle,rounded corners=2pt,xshift=3cm]  {$\true$};
	
	\node (a4) [below left of=a3,draw,rectangle,rounded corners=2pt,xshift=-1cm] {\playaction{r!=0\&driveUser=0->r'=0\&gc'=true}};
	\node (good2) [right = of a4,draw,rectangle,rounded corners=2pt,xshift=1cm]  {$\true$};

	\node (a5) [below right of=a4,draw,rectangle,rounded corners=2pt,xshift=4cm,yshift=-0.1cm] {\playaction{s2=0\&r2=0->r2'=2}};
	\node (good3) [left =of a5,draw,rectangle,rounded corners=2pt,xshift=-3cm]  {$\true$};

	\node (a6) [below left of=a5,draw,rectangle,rounded corners=2pt,xshift=-6cm] {\playaction{s2=0\&r2=2->s2'=1}};
	\node (good4) [right = of a6,draw,rectangle,rounded corners=2pt,xshift=4cm]  {$\true$};

	\node (a7) [below right of=a6,draw,rectangle,rounded corners=2pt,xshift=5.5cm,yshift=-0.05cm] {\playaction{any synchronized action}};
	\node (good5) [left = of a7,draw,rectangle,rounded corners=2pt,xshift=-3cm]  {$\true$};
	
	\node (a8) [below left of=a7,draw,rectangle,rounded corners=2pt,xshift=-4.5cm] {\playaction{u1\_request\_comm}};
	\node (bad2) [right =of a8,draw,rectangle,rounded corners=2pt,xshift=3.5cm]  {$\false$};
	
	\node (bad3) [below left of=a8,draw,rectangle,rounded corners=2pt,xshift=-.5cm] {$\true$};
	\node (good6) [below right of=a8,draw,rectangle,rounded corners=2pt,xshift=2.5cm] {$\false$};
	
	\node (u11) [right =5pt of a1,draw,circle] {$1$};
	\node (u12) [right =5pt of a2,draw,circle] {$1$};
	
	\node (u21) [right =5pt of a5,draw,circle] {$2$};
	\node (u22) [left =5pt of a6,draw,circle] {$2$};
	
	\draw [-] (a1) to (bad1);
	\draw [-,dashed] (a1) to (a2);
	
	\draw [-] (a2) to (good1);
	\draw [-,dashed] (a2) to (a3);	

	\draw [-] (a3) to (a4);
	\draw [-,dashed] (a3) to (good2);
	
	\draw [-] (a4) to (good3);
	\draw [-,dashed] (a4) to (a5);
	
	\draw [-] (a5) to (good4);
	\draw [-,dashed] (a5) to (a6);	
	
	\draw [-] (a6) to (good5);
	\draw [-,dashed] (a6) to (a7);	
	
	\draw [-] (a7) to (a8);
	\draw [-,dashed] (a7) to (bad2);
	
	\draw [-] (a8) to (bad3);
	\draw [-,dashed] (a8) to (good6);
	
\end{tikzpicture}
}

%\end{document}
	\caption{A decision tree for \texttt{mer}}
	\label{fig:mertree}
	\myspace
\end{figure}
\myspace\myspace
\end{center}

%Conflicts in IP address are detected by broadcast messages after choosing an appropriate response to answers from other network participants (environment hosts).
%Once a user has waited a sufficient enough time for responses that would force him to choose a new IP address, it sticks to its current address and does not change it again.
%The property we model check is that there are two network participants end up having the same IP address.
%The probability of reaching such a target state is roughly $0.008$.
%
%\begin{minipage}{0.7\textwidth}
%\begin{figure}
    %\resizebox{0.75\linewidth}{!}{
	%\begin{subfigure}{\textwidth}
	%\input{figures/zeroconf_tree_2.tex}
	%%\caption{A decision tree for {\texttt zeroconf}}
	%\label{fig:zeroconftree}
	%\end{subfigure}
	%}
%%\end{minipage}
%%\begin{minipage}{0.3\textwidth}
%\resizebox{0.2\linewidth}{!}{
	%\begin{subfigure}{0.2\textwidth}
	%\input{figures/firewire_tree.tex}
	%%\caption{Trivial tree}
	%\label{fig:firewiretree}
	%\end{subfigure}
	%}
	%\caption{Decision trees for \texttt{zeroconf} (left) and a trivial one for \texttt{firewire} (right).}
%\end{figure}
%%\end{minipage}

\afterpage{\clearpage}

\section{Conclusion}
In this work we presented a new approach to represent strategies in MDPs in a 
succinct and comprehensible way. 
We exploited machine learning methods to achieve our goals.
Interesting directions of future works are to investigate whether other machine 
learning methods can be integrated with our approach, and to extend our approach
from reachability objectives to other objectives (such as long-run average and discounted-sum).
% \todo{fill in 4 lines}

\newpage

\bibliographystyle{alpha}
\bibliography{refs,related-work,bib,diss}

\iffalse
\section*{Sketch of structure (discussed in Sydney)}

1. Intro [2/12]

2. Definitions: MDPs, Prob measure, strategies, properties, etc. [1 1/2]

3. Data-structures for Strategies [2 1/2 ]
 A) Explicit
 B) BDD
 C) Decision trees

Advantages of decision trees: other only one strategy, multiple suboptimal, and others.
Illustration with examples.

4. Theoretical Bounds [1 1/2]
 Formulate the theoretical question and any lower bound.

5. Practical Algorithms [2 1/2]
 A) Small state space (start from complete strategies)
 
 B) Large state space (start with ML learned strategies)

Illustration with Examples

6. Experimental Results [3-4] 
A) For 6A
B) For 6B
Along with illustrations.

7. Conclusion and Future Work [1/2]
\fi

\end{document}